\documentclass[journal]{IEEEtran}
\usepackage{stmaryrd}
\usepackage{amsmath}
\usepackage[section]{placeins}
\usepackage{amssymb}
\usepackage{psfrag}
\usepackage{epstopdf}
\usepackage{todonotes}
\usepackage{graphicx}
\usepackage{algorithm}
\usepackage{svg}
\usepackage{mathtools}
\usepackage{hyperref}
\usepackage{booktabs}
\usepackage{hyphenat}
\usepackage{multirow}
\usepackage{booktabs}
\usepackage{textcomp}
\usepackage[utf8]{inputenc}
\usepackage{array,booktabs,calc}
\usepackage{soul}
\usepackage[caption=false]{subfig}
\usepackage{float}

\begin{document}
\title{{Towards Auto-Modeling of Formal Verification for NextG Protocols: A Multimodal cross- and self-attention Large Language Model Approach}}

\author{\IEEEauthorblockN{Jingda Yang\IEEEauthorrefmark{1} ~~
Ying Wang\IEEEauthorrefmark{1} ~~}\\
\IEEEauthorblockA{\IEEEauthorrefmark{1}Stevens Institute of Technology, Hoboken, NJ \\
Email: \IEEEauthorrefmark{1}\{jyang76, ywang6\}@stevens.edu}}

\maketitle

\begin{abstract}

This paper introduces Auto-modeling of Formal Verification with Real-world Prompting for 5G and NextG protocols (AVRE), a novel system designed for the formal verification of Next Generation (NextG) communication protocols, addressing the increasing complexity and scalability challenges in network protocol design and verification. Utilizing Large Language Models (LLMs), AVRE transforms protocol descriptions into dependency graphs and formal models, efficiently resolving ambiguities and capturing design intent. The system integrates a transformer model with LLMs to autonomously establish quantifiable dependency relationships through cross- and self-attention mechanisms. Enhanced by iterative feedback from the HyFuzz experimental platform, AVRE significantly advances the accuracy and relevance of formal verification in complex communication protocols, offering a groundbreaking approach to validating sophisticated communication systems. We compare CAL's performance with state-of-the-art LLM-based models and traditional time sequence models, demonstrating its superiority in accuracy and robustness, achieving an accuracy of 95.94\% and an AUC of 0.98. This NLP-based approach enables, for the first time, the creation of exploits directly from design documents, making remarkable progress in scalable system verification and validation.

\end{abstract}



\section{Introduction}

The third generation partnership project (3GPP) published its Release 17 (Rel-17) specifications at the end of the first quarter of 2022~\cite{3gppRel17_2022}. Since then, parallel and subsequent releases have been rolled out to enhance and address new and unfulfilled requirements from previous releases. It ensures ongoing innovation and improvement in mobile communication technologies, aligning with the evolving needs and advancements in the field. 3GPP protocols consist of a multitude of technical specifications and documents that cover various aspects of mobile communication networks, including radio access, core network, and service capabilities. 3GPP protocols, encompassing a wide array of technical specifications, are extensive, with documents like the Radio Resource Control (RRC) in Release 17 spanning over a thousand pages~\cite{3GPP_TS_38331_v1700}. The extensive nature of these documents, coupled with their distributed security protocols, renders manual verification and testing both time-intensive and susceptible to error. This complexity escalates in future network generations, exacerbating the risk of zero-day vulnerabilities.

Furthermore, the incorporation of O-RAN and network function virtualization introduces additional layers of complexity and potential attack vectors through cloud APIs into 5G and future G infrastructure~\cite{cloudapi}. These advancements offer enhanced functionality but expose operators to novel security challenges, especially in the relatively uncharted territory of cloud security. This complexity, compounded by the involvement of various entities in development, heightens the risk of security breaches, particularly through misconfigured containers and exposed APIs.

Thus, in 5G and future G systems and networks, logical attacks exploiting protocol logic errors represent a significant vulnerability category. These attacks are challenging to detect due to the scale and complexity of the network systems. Formal verification encompasses a broad range of techniques used to prove or disprove the correctness of algorithms, protocols, systems, or software with respect to certain formal specifications. It also involves mathematical analysis to ensure that a system behaves as intended. Utilizing the Dolev-Yao (DY) formal attack model~\cite{dolev1983security, cervesato2001dolev}, one of the widely adopted methodologies, formal verification has demonstrated its effectiveness in identifying flaws in infrastructure and communication protocols. However, this approach is limited to abstract specifications, and, as of now, the full automation of protocol verification and validation of their implementations remains beyond the reach of current technologies~\cite{ammann2024dy, rakotonirina2024decision}. 


With the recent advancements in Large Language Models (LLMs), an intriguing question emerges: Can LLMs contribute to accelerating the design and verification of large-scale protocols, and can they be effectively integrated with system validation processes that involve existing implementations?


To answer this question, two crucial conditions must be addressed: clarifying ambiguities and capturing design intentions. The first condition involves resolving the conflict between the ambiguity inherent in natural language processing and the need for explicitness in formal verification modeling. Resolving this conflict is essential to demonstrate the potential to provide measurable and verifiable trustworthiness. The second condition requires differentiating intentional relationships or dependencies from unintentional ones identified by LLMs in targeting protocol designs. The differentiation is crucial in transforming the design-intended relationships into mathematical and logical expressions in formal verification.

In the past several months, both academic and industry sectors have increasingly focused on applying LLMs in the realm of formal verification. Two major areas of this application are using prompt engineering and LLMs for hardware assertion~\cite{kande2023llm,cosler2023nl2spec,orenes2023using} and software system Bounded Model Checking (BMC)~\cite{kande2023llm, cosler2023nl2spec}. To our knowledge, no published work yet explores the use of LLMs in verifying communication-related protocols and specifications, especially for the large and complex protocols of 3GPP releases. One of the reasons is due to a common challenge in existing research is capturing trustworthy design intent in a set of assertions for use in formal verification or testing-based checking. This challenge intensifies as system distribution and complexity increase in the case of 5G and the future G, particularly when intertwined with a broad spectrum of various usage scenarios and verticals. The iterative prompts to the LLMs has limitations due to the complex and broad dependencies among identifiers, commands, and properties.

Further more, our previous work on non-LLM-based NLP approach for 5G and other communication protocols \cite{yuan2023ambiguity} marked a significant shift from manual to automated, accuracy-focused analysis in translating natural language oriented protocols into formal models. While it revealed limitations in handling complex semantic relationships with strong contextual control, it also highlighted the potential and direction for LLM-based NLP in formal modeling. We found this approach becomes particularly promising in detecting design intentions with greater intelligence and enhancing trustworthiness through its connection to learn from real-world experiments~\cite{hyfuzz2023}.   


In this paper, we introduce a novel approach, named Auto-modeling of Formal Verification with Real-world Prompting for 5G and NextG protocols (AVRE), which addresses the need for scalable formal verification in the domains of 5G and NextG. AVRE uses LLMs to clarify ambiguities and capture design intent, transforming protocol descriptions into dependency graphs and formal models. Our method differs from existing techniques by integrating a transformer model with the LLM, allowing quantifiable dependency relationships to be generated under supervision and enabling transformative learning without human involvement. This system, enhanced by iterative feedback from an experimental platform: HyFuzz~\cite{hyfuzz2023}, fills a research gap by combining experience-based and logical dependency analyses in protocol documentation, thereby significantly improving the accuracy and relevance of formal verification in complex communication protocols.

\subsection{Related Work}\label{related_work}

\textbf{Converting Informal Natural Language System Designs and Protocols to Formal Description}: Approaches to transforming natural language descriptions into formal models have seen considerable advancements, evolving through the introduction of diverse methodologies over the years. A decade ago, Drechsler et al. \cite{drechsler2012formal} proposed a paradigm that incorporated the Formal Specification Level (FSL), adeptly bridging the gap between informal textbook specifications and formal Electronic System Level (ESL) interpretations. Subsequently, Banarescu et al. \cite{banarescu2013abstract} proposed a hybrid methodology that converted linguistic expressions into formal paradigms by merging symbolic and statistical techniques. With the advent of deep learning, Dong and Lapata \cite{dong2016language} employed neural networks to convert natural language instructions into executable codes. This work was further enhanced by the contributions of Reddy et al. \cite{reddy2019coqa} in 2019, who focused on semantic parsing, utilizing denotations to transform complex linguistic structures into formalized notations. Meanwhile, the application of regular expressions has been identified as a viable means to extract formal specifications from natural language narratives, providing advanced components in deep learning frameworks \cite{hahn2022formal}. Despite these developments, models based on these previous methodologies have achieved an accuracy threshold of approximately 90\%, which is inadequate for ensuring complete recall. This poses a challenge in the precise conversion of natural language protocols into formal formulations. 

\textbf{LLMs Based Formal Verification}: 
LLMs have demonstrated impressive reasoning and assertion capabilities for formal verification \cite{ cosler2023nl2spec, saparov2022language, srikumar2023fast}. Research in \cite{kande2023llm, cosler2023nl2spec} has explored using LLMs to generate temporal logic specifications and assertions from unstructured natural languages. Meanwhile, studies in \cite{tihanyi2023formai, charalambous2023new} focus on leveraging LLMs to enhance BMC for identifying software vulnerabilities and deriving counterexamples. In \cite{orenes2023using}, the authors trained GPT-4 to generate correct SystemVerilog Assertions (SVA) through iterative prompt refinement with rules. However, it remains unclear how these models derive answers and whether they rely on simple heuristics rather than a generated chain-of-thought \cite{saparov2022language}. The current state of the art prioritizes producing formal specifications and properties quickly, albeit with slight inaccuracies, over generating perfect specifications or correctness statements \cite{srikumar2023fast}. The non-transparency related to LLM heuristics leads to a large number of irrelevant dependencies, resulting in low precision in dependency classification. To address this, our experimental platform connects to guide and refine the dependency graph range.

\textbf{Prompting Limitations in LLM Enabled Formal Verification}: 
Furthermore, the majority of existing work relies on prompt engineering. LLM-integrated applications blur the line between data and instructions \cite{abdelnabi2023not}. LLMs can produce non-deterministic outputs, potentially yielding different results for the same prompt. This variability poses a potential threat to the validity of scientific conclusions unless researchers adapt their methods to account for it in their empirical analyses \cite{ouyang2023llm}. The adoption of prompting methods introduces challenges in iterative formal verification without human involvement. The randomness in LLMs is influenced by the sampling methods used during text generation, such as top-k sampling or nucleus sampling \cite{krishna2022rankgen}, limiting its application in classifiers or deterministic types of applications. To address the non-determinism and iterative formal verification, distinct from current formal verification methods that utilize prompt engineering, we designed an open-access LLM, integrated with a transformer model, to achieve supervised dependency."

\textbf{Digital Engineering Aided Formal Guided System Validation}: 
In the field of integrated Design Validation, there has been recent research progress in combining formal verification with simulation, resulting in a practical validation engine with reasonable run-time \cite{li2022digital}. Experimental work in the context of 5G has gained significant attention over the past few years, shifting from the simulation-driven research used in previous mobile network generations to system implementation prototyping \cite{wang2021development}. In our previous work, we explored a Formal-guided Fuzzing testing approach \cite{yang2023formal, yang2023formalampli} to bridge design verification and system validation. This approach complements the scalability limitations of formal verification and addresses the impacts of detected vulnerabilities. We have introduced a fuzzing digital twin \cite{dauphinais2023automated} to provide an open and automated platform for systematically that enables an autonomous detection of vulnerabilities and unintended emergent behaviors in 5G infrastructures. 
However, a constraint of this initial approach is heavy reliance on expert insights to identify and articulate formal relationships.

\subsection{Contribution}\label{contribution}

The main original contributions of this work are summarised as follows: 
\begin{itemize}
    \item The research introduces a novel approach that utilizes LLMs to address critical challenges in formal verification, specifically in clarifying ambiguities and capturing design intent. This approach combines experience-based prior-probability distribution with logical dependency analysis in protocol documentation, and leverage experimental-based posterior-probability to enhances the accuracy and relevance of the dependency graph by the cross-attention-based LLM (CAL) model.

    \item The paper presents a new systematic solution named AVRE, which equipped with CAL, a continuously-learning, cross-attention-based LLMs. CAL is desined to interpret and transform protocol descriptions into dependency graphs, which can then be converted into formal models.  CAL is enhanced by incorporating iterative feedback from an experimental platform, HyFuzz, to focuse on refining the capture of intentions and resolution of ambiguities.
    
    \item In CAL, we introduce a novel multi-session detection method that bypasses traditional token count barriers. By segmenting comprehensive protocols into manageable sections and concurrently processing them, it can output detailed and quality analysis without token constraints. CAL incorporates refined cross-attention mechanisms instead of prompting based scheme, achieves high efficiency to identify and interpret complex formal relationships with enhanced accuracy and insight.

    
    
    \item We have developed a scalable cross sessions dependency graph that supports the hierarchy of formal analysis, facilitating the revelation of in-depth relationships embedded within protocols to systematically pinpoint and counteract vulnerabilities.  
    
    
\end{itemize}





The rest of our paper is organized as follows. Section \ref{related_work} listed the existing literature in the field of LLMs, dependency graph, formal verification, and experimental based validation. We introduce our proposed EVRA and show the architecture of the system in Section \ref{system_model}, followed by the the detailed description of the system in Section \ref{protocol_description} and Section \ref{model_description}. We provide the performance analysis and in Section \ref{result}. Lastly, we concluse our work with future research directions in Section \ref{conclusion}.

\section{System Overview}\label{system_model}

\begin{figure*}[h!]
\centering
    \includegraphics[width=0.9\textwidth]{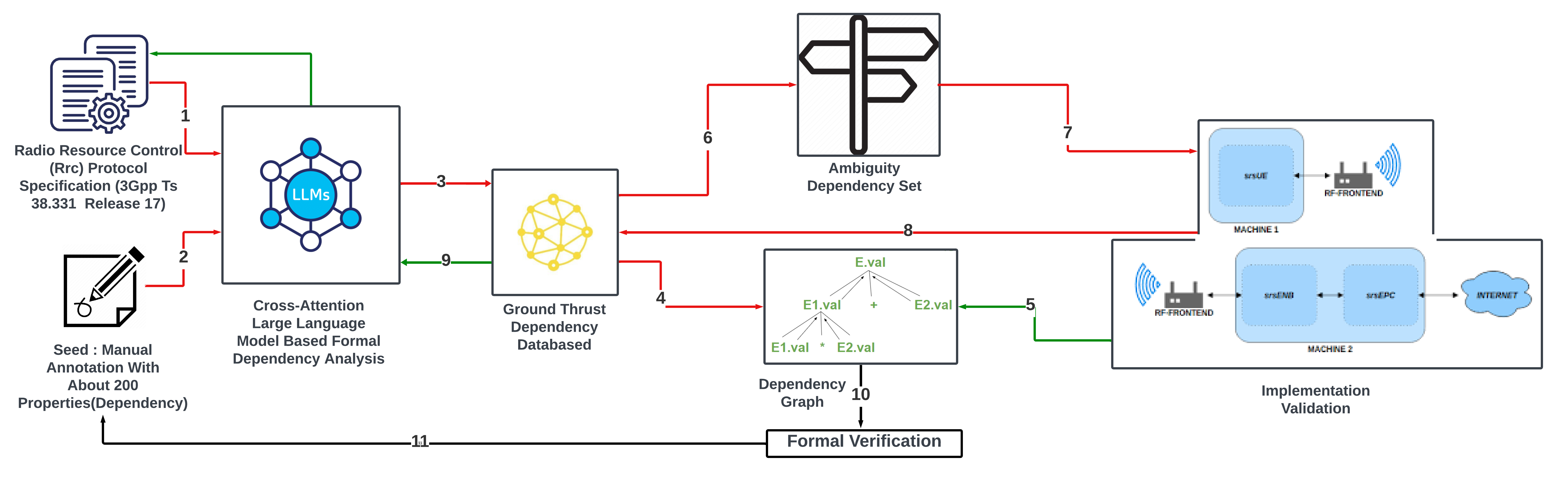}
    \caption{System Overview of Auto-Modeling and Trustworthy for Formal Verification and Validation in 5G and NextG Security Protocols. Red line shows the process from informal system protocols to dependency graph and formal expression, and green line shows the formal guided fuzz testing feedback to the CAL and refine the results}
    \label{fig:system_model}
\end{figure*}

In addressing the critical need for scalable formal verification in the rapidly evolving domains of 5G and NextG protocols and specifications, in this paper, we present a pioneering approach that utilizes LLMs to overcome key challenges of clarifying ambiguities and capturing design intent. We present a novel systematic solution AVER as shown in Fig. \ref{fig:system_model} : Auto-modeling of Formal Verification with Real-world Prompting for 5G and NextG protocols.

At the heart of AVRE is a continuously-learning, CAL designed to interpret and transform protocol descriptions into dependency graphs, which are subsequently converted into formal models. Differing from the current state of the art in using LLM prompting for formal verification, our approach integrates the transformer model with the LLM model, enabling it to generate a quantifiable dependency relationship under supervision. This facilitates the transformative learning of the LLM without human involvement, meeting the requirements for formal explicitness. This system is further enhanced by incorporating iterative feedback from our experimental platform: HyFuzz~\cite{hyfuzz2023}, aimed at refining intention capture and ambiguity resolution.

CAL is trained on labeled identifiers (step 1 and step 2) and formal properties, designed to predict formal properties (step 3) with a more controlled environments. The design of cross-attention mechanisms utilization the LLM model improve the capability of the model in complex protocol. 

The experimental platform serves as two purpose: intention and the trustworthy. As shown in step 4 and step 5 in Fig. \ref{fig:system_model}, it provides relevancy and design intention information , which is used with the dependency relationships detected by CAL together to form the dependency graph.  For trustworthy, as shown in step 6 and step 7, the low confident prediction is set to the experimental platform to return evidence that could be added to ground truth dependence database. With multiple iterations, a robust dependency graph will be generated to sent to formal method verification. The detected vulnerability will be further used in protocol fortification.



The design of the system AVRE uniquely fills a gap in the existing research landscape by integrating experience-based prior-probability distribution with logical dependency analysis in protocol documentation using LLMs. Moreover, by leveraging experimental-based posterior-probability through real-world prompting, AVRE establishes an iterative learning loop, significantly enhancing the accuracy and relevance of the intention of the CAL model. This advancement marks a pivotal contribution to the field, offering new directions in the formal verification of complex communication protocols.

\section{Protocol Analysis and Data Pre-Process and Annotation}\label{protocol_description}
\subsection{Protocol Analysis and Formal Properties Definition}

In this paper, we use 3GPP Release 17 Radio Resource Control \cite{ETSI2022} as an example to illustrate AVRE. We identify two fundamental elements: procedures and identifiers. Procedures are defined as sequences of actions and interactions among different entities. Identifiers in the protocols are specific labels or names used to uniquely recognize various elements within mobile telecommunications networks. For instance, IMSI (International Mobile Subscriber Identity) is a unique number associated with all cellular networks, used primarily for identifying individual subscribers for billing and identification purposes. IMEI (International Mobile Equipment Identity) is defined as a unique number to identify mobile devices, primarily used for identifying the device itself, rather than the subscriber \cite{ETSI2022}. The procedures define the structure of a set of identifiers that are connected dependently. The dependent relationship between the identifiers, framed by the structure defined in the procedures, generates the dependency graph, which is then converted for formal analysis. As shown in Table. \ref{table:rrc_setup_request}, while the extraction of entities from protocol identifiers is straightforward, the classification of their dependency is a problem that CAL targets. However, the procedures include intricate interactions and designs, with unspecified space for various vendors to implement based on their existing infrastructures and devices. In this paper, we innovatively connect to real-world or simulated experimental platforms to aid in procedure formulation.



\begin{table}[h!]
\centering
\caption{RRCSetupRequest-IEs field descriptions}
\begin{tabular}{|p{0.27\columnwidth}|p{0.6\columnwidth}|}
\hline
\textbf{Identifier} & \textbf{Description} \\
\hline
establishmentCause & Provides the establishment cause for the \texttt{RRCSetupRequest} in accordance with the information received from upper layers. gNB is not expected to reject an \texttt{RRCSetupRequest} due to unknown cause value being used by the UE. \\
\hline
ue-Identity & UE identity included to facilitate contention resolution by lower layers. \\
\hline
\end{tabular}
\label{table:rrc_setup_request}
\end{table}

Considering confidentiality, integrity, and availability (CIA) triad which is widely accepted security model, along with the Dolev-Yao model for communication protol formal verification, we have proposed four essential properties to describe the dependency among identifiers \cite{yang2023formal}. These properties, each addressing a distinct facet of security augmentation within the specifications, are defined as follows:
    \begin{enumerate}
        \item \textbf{Confidentiality}: the capability of source to prevent private information from leakage of the destination. The source should be the selected encryption algorithm or the key of encryption algorithm, and the destination should be command or specific identifier, which are confidentiality protected by source. 
        
        \item \textbf{Integrity} : the capability of source to keep the destination information unmodified. Similar to confidentiality, the source should be the selected integrity algorithm or the key of integrity algorithm, and the destination should be the command or specific identifier, which are integrity protected by source.
        \item \textbf{Authentication}: the ability of source to help User Equipment (UE) or gNodeB (gNB) to identify where and when the destination was sent from. The source should be the distinctive identifier, which can uniquely represent a entity (like gNB) or a communication session (like UE id).
        \item \textbf{Accounting}: the source of accounting relationship is the counting identifier which can sequentially track and distinguish each transmitted command. The destination should be the sequence-protected command.
        \item \textbf{Include}: the destination identifier is included in the source identifier.
        \item \textbf{Generate}: the destination identifier is generated by the source identifier.
    \end{enumerate}


\subsection{Data Preparation}\label{sec:data_prep}

In our study, we annotated dependencies (formal properties) of identifiers in the 5G Radio Resource Control (RRC) protocol \cite{ETSI2022}, focusing on Sections 5.3.1 to 5.3.5. From these sections, approximately 16,428 samples were annotated, including source, destination, and dependency relationships. Out of these, 1,218 samples were identified by domain experts who analyzed the documents to determine relationships. The remaining samples consist of source and destination pairs lacking a dependency relationship. In our proposed model, we treat this as a classification problem for all potential source-destination pairs, encompassing both relevant and non-relevant ones. Consequently, the large number of source and destination pairs results in highly imbalanced data. 

\begin{figure}[h!]
    \includegraphics[width=0.5\textwidth]{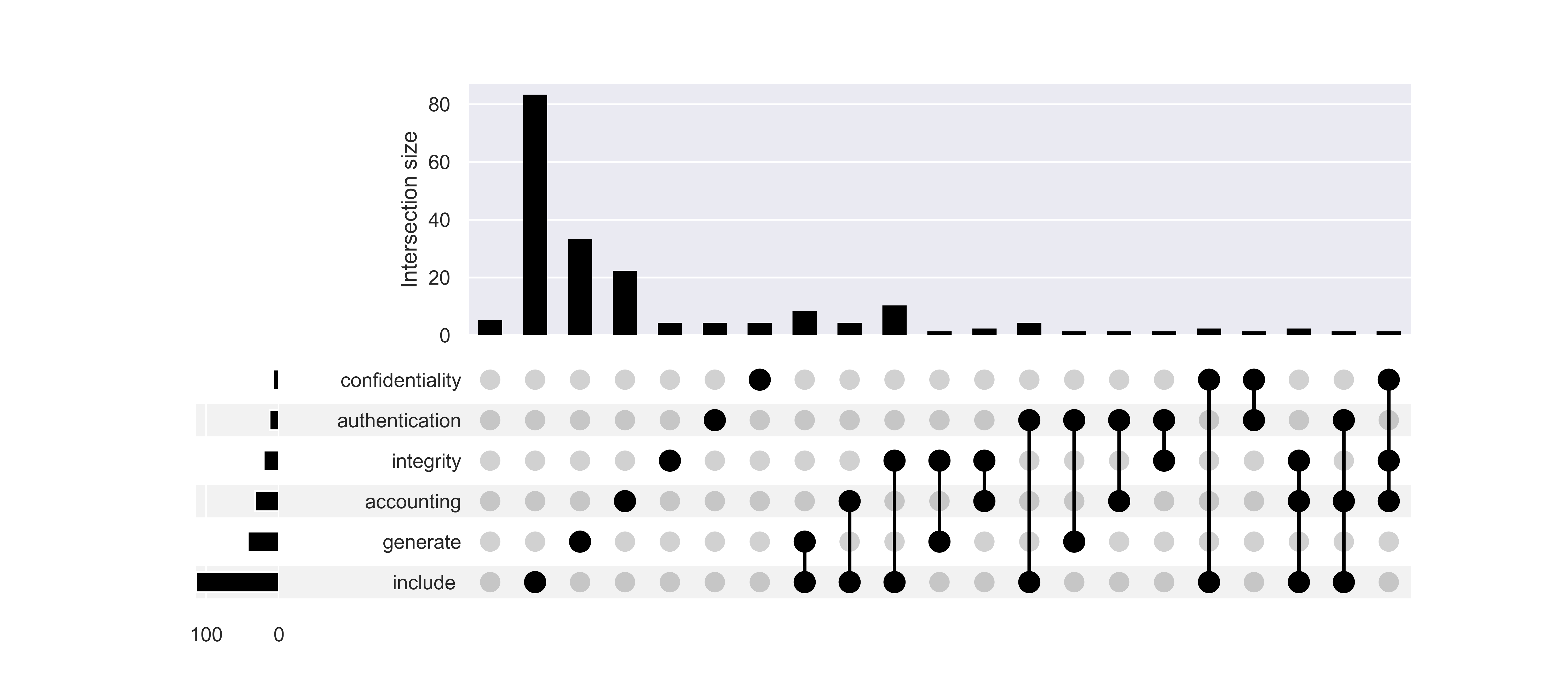}
    \caption{Formal Properties Statistical Counts and Intersections}
    \label{fig:upset}
\end{figure}

In our analysis of the annotated data samples, we have identified an imbalance in the distribution between positive and negative samples. To address this issue and improve the model's performance, we propose employing a weight-balanced binary cross-entropy loss. Fig. \ref{fig:upset} illustrates the size of each property type and the intersection sizes among them. It is clear from Fig. \ref{fig:upset} that 'include' has the largest number of detections. Furthermore, the intersections of 'include' with 'account' and 'integrity' represent the two largest intersection counts, indicating a strong correlation between these properties. It is also important to note the significant presence of 'integrity' in multiple intersections, which is consistent with the importance of integrity in RRC protocols 3GPP in release 17.

\section{Methodology}\label{model_description}
\subsection{Building Multimodal cross- and self-attention Large
Language Model}

\begin{figure}[h!]
    \includegraphics[width=0.5\textwidth]{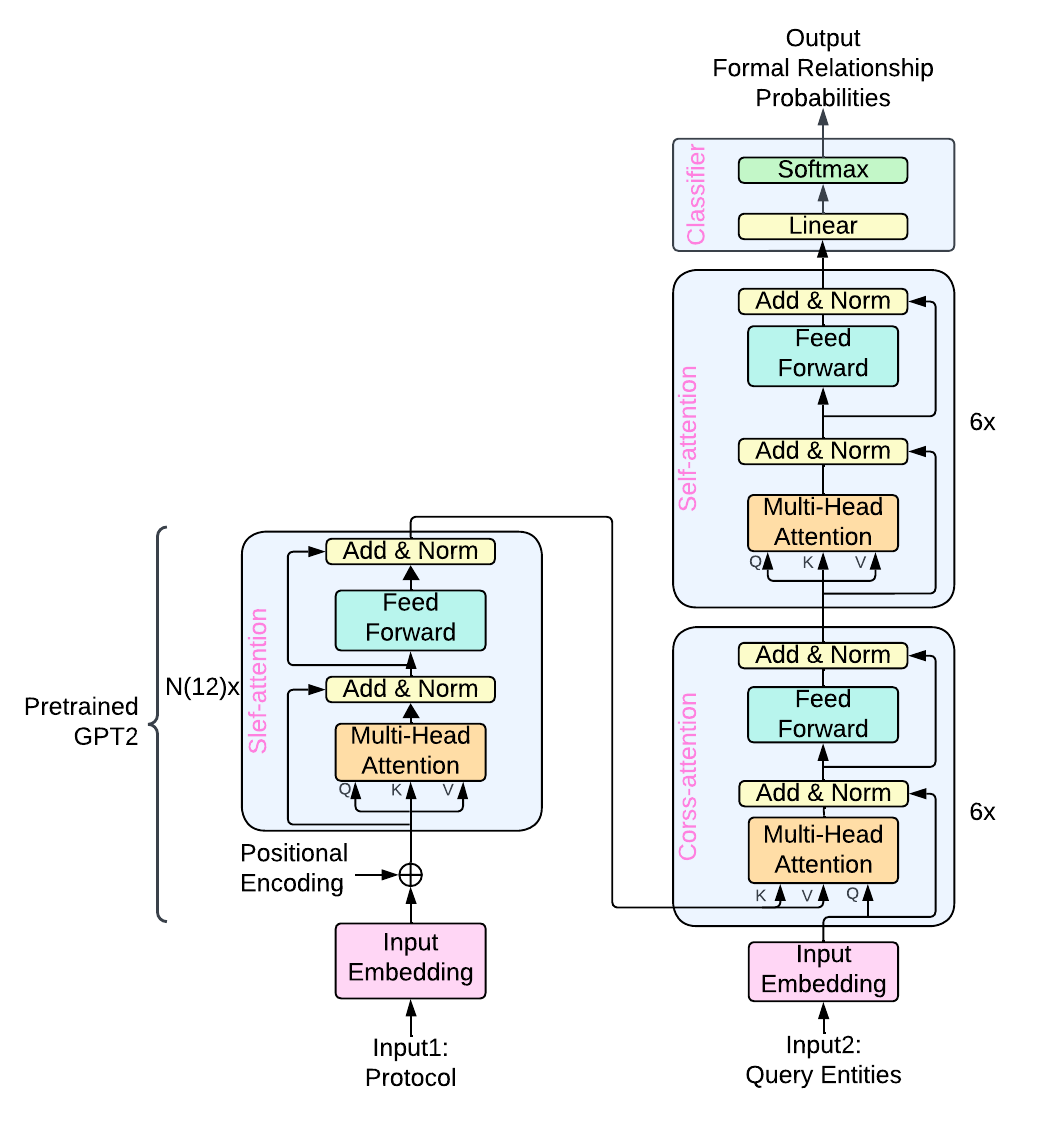}
    \caption{Attention Driven Formal Identifier and Property Abstraction Model }
    \label{fig:model_view}
\end{figure}

In this work, we proposed CAL,  an LLM embedded with cross- and self-attention model, as delineated in Fig.~\ref{fig:model_view}, considering both the contextual information from the original protocols and learns the defined dependency relationships among identifiers. CAL model employs a pretrained LLM (GPT-2), which consists of $N$ transformer layers, to extract hidden insights from protocol descriptions. Here, considering the scalability of training and the performance accuracy, we select $N = 12$.

We incorporate cross-attention mechanisms to discern the relationships between the extracted latent information and query entities~\cite{46201}. To further enrich the contextual understanding, we deploy self-attention frameworks that evaluate inter-relations among all positions from the preceding stage, guided by weighted considerations. In the final stage of the presented model, a linear classifier is implemented to infer probabilities associated with distinct formal attributes.

Self-attention generates contextual representations for a single sequence by computing weighted averages of all tokens, while cross-attention evaluates interdependent contextual relationships between the query sequence and the context sequence in transformer models~\cite{vaswani2017attention,devlin2018bert,brown2020language}. In the proposed CAL model for converting complex contextual protocols into explicit dependency graphs, self-attention serves as the mechanism to understand the context within the protocols, while cross-attention understands and recognizes the relationships across the identifier sequences.

Self-attention calculates a weighted average of feature representations, where the weight represents the similarity score between pairs of feature representations. As defined in ~\cite{vaswani2017attention}, an input sequence of $n$ tokens of dimensions $d$, $X\in R^{n\times d}$, is extracted by three projection matrices $W_Q\in R^{d\times d_q}$, $W_K\in R^{d\times d_k}$ and $W_V\in R^{d\times d_v}$ ($d_q=d_k$). The input consists of queries and keys with dimension $d_k$, and values with dimension $d_v$. Three different feature representations $Q = XW_Q,K = XW_K,$ and $V = XW_V$, where $Q\in R^{n\times d_q}$, $K\in R^{n\times d_k}$ and $V\in R^{n\times d_v}$. Intuitively, $Q$, $K$ and $V$ are separately regarded as query, key and value. Normalized dot-product of query and key is used to represent the attention score $A\in R^{n\times n}$ of each query and paired keys as:
\begin{equation}
    A = softmax(\frac{QK^T}{\sqrt{d_q}})
\end{equation}
where division by $\sqrt{d_q}$ normalizes the dot-product of $Q$ and $K^T$, ensuring that the distribution of the dot-product aligns with expectation $\mathrm{E}=0$ and variance $\mathrm{Var}=1$. Ultimately, attention value is calculated as follow, 
\begin{equation}
    \mbox{Self-attention}(Q,K,V) = softmax(\frac{QK^T}{\sqrt{d_q}})V=AV
\end{equation}
where output $\mbox{Self-attention}(Q,K,V) \in R^{n\times d_v}$ is the weighted average of extracted features. Normally, a residual linear feed forward layer is incorporated to further distill the acquired knowledge represented by the weighted average features.

Utilizing the self-attention mechanism, cross-attention takes two input sequences with different sequence length $X_1\in R^{n_1\times d}$ and $X_2\in R^{n_2\times d}$ as inputs. Subsequently, $Q = X_1W_Q,K = X_2W_K,$ and $V = X_2W_V$ are derived from $X_1$ and $X_2$ respectively and the dimension of attention value can be succinctly expressed as $R^{n_1\times d_v}$.

\subsection{Balanced Loss Function}

In Section \ref{sec:data_prep}, we observed the imbalance in the data distribution of relationships amongst entities, and is evident in Table~\ref{tab:class_distribution}. In response to this challenge, we adopt a weight-balanced binary cross-entropy loss, articulated in Equation~\ref{equation:weight_BCE}. In this equation, $N$ signifies the total number of cases, $y_i$ represents the ground truth, $p(y_i)$ is the predicted probability, and $n_{(y_i)}$ denotes the proportion of class $y_i$ within all cases. Utilizing the inverse class ratio amplifies the impact of the underrepresented class while tempering the dominance of the majority class, offering a balanced approach to mitigate the skewed data distribution.
\begin{table}[ht]
\centering
\caption{Imbalanced Nature of Formal Property Data}
\begin{tabular}{|l|c|r||l|c|r|}
\hline
\textbf{Relationship} & \textbf{Class} & \textbf{Count} & \textbf{Relationship} & \textbf{Class} & \textbf{Count} \\
\hline
Confidentiality & 0 & 16421 & Accounting & 0 & 16399 \\
\hline

Confidentiality & 1 & 7 & Accounting & 1 & 29 \\
\hline

Integrity & 0 & 16408 & Include & 0 & 16329 \\
\hline

Integrity & 1 & 20 & Include & 1 & 99 \\
\hline

Authentication & 0 & 16417 & Generate & 0 & 16389 \\
\hline

Authentication & 1 & 11 & Generate & 1 & 39 \\
\hline
\end{tabular}
\label{tab:class_distribution}
\end{table}

\begin{equation}
\label{equation:weight_BCE}
\begin{gathered}
    L_i = -\left[y_i\cdot log(p(y_i)) + (1-y_i) \cdot log(1-p(y_i)) \right] \\
    L = \frac{1}{N}\sum_{i=1}^{N}L_i   \cdot (1- \frac{n_{(yi)}}{N})
    \end{gathered}
\end{equation}

\subsection{Connection to Experimental Platform}
Inspired by the iterative prompting to the LLM, the experimental platform serves as the prompting server in real world. In order to clarify the ambiguity in the LLM model, the counterpart experiments configuration could be auto-generated and performed, in which the results generated is feedback to the LLM to improve the trustworthiness. Through the digital engineering module, which facilitates connections to available real-world execution platforms or digital twins, the process effectively bridges the gap between design intentions and real-world operations for mission-critical infrastructures. 

\begin{figure}
    \centering
    \colorbox{yellow!20}{%
    \includegraphics[width=0.5\textwidth]{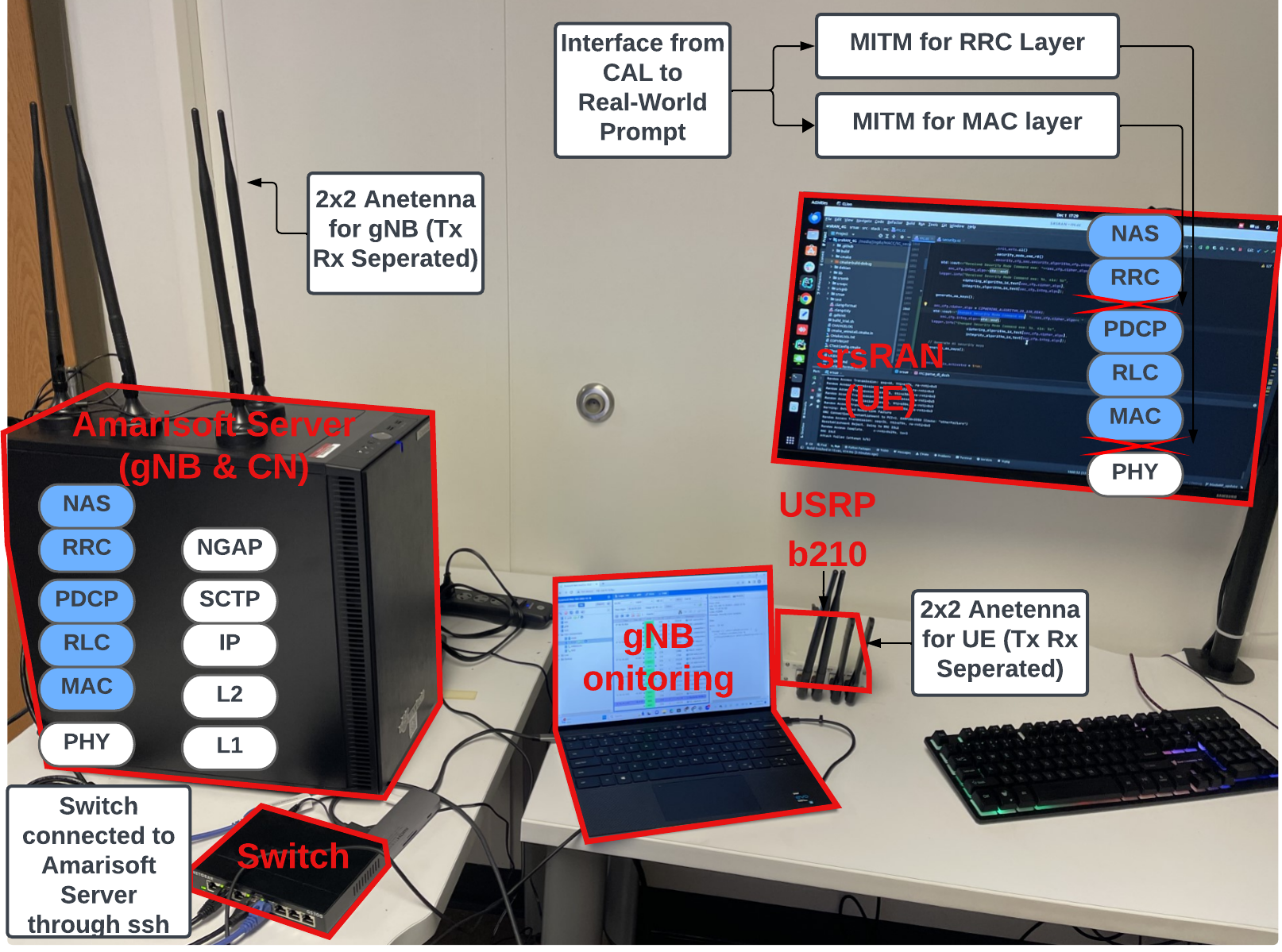}}
    \caption{Virtual and Over the Air (OTA) Mode Experimental of HyFuzz\cite{hyfuzz2023}}
    \label{fig:devices}
\end{figure}

The experimental platform, HyFuzz, employs a hybrid system model \cite{hyfuzz2023} that incorporates two distinct platforms: a ZeroMQ virtual model and an Over the Air (OTA) physical model. In addition to running 5G tests in various scenarios, the uniqueness of the HyFuzz platform is its ability to serve as relay nodes to perform various fuzzing tests. The relay node is capable of modifying and permutating commands and accessing messages exchanged between the UE and the gNB. HyFuzz facilitates multi-step Man-In-The-Middle (MITM) attacks, enabling the identification and analysis of vulnerabilities and the detection of various prompts due to the uncertainty of CAL model outputs and providing an accurate real-world ground truth. An overview of the HyFuzz is shown in Fig. \ref{fig:devices}. 

As illustrated in Fig. \ref{system_model}, the experimental platform is seamlessly integrated with the CAL model, supporting it from two key aspects: design intention and trustworthiness. Regarding design intention, the platform supplies relevant information and design intents to the CAL model, enabling it to continue training based on novel evidence of dependency relationships, thus effectively constructing an accurate dependency graph. In terms of trustworthiness, low-confidence predictions are referred back to the experimental platform for further validation, potentially contributing new data to the ground truth dependency database. This iterative process aims to create a robust dependency graph, which will be utilized for formal method verification. Detected vulnerabilities will then be leveraged in protocol fortification."

\section{System Performance Assessment}\label{result}
\subsection{CAL Experiment Setting}
\label{experiment_setting}

Considering the openness and accessibility of the LLMs, in this paper, we select the pre-trained GPT-2 model with 12 transformer layers and 12 attention heads as embedded pre-trained LLM included in CAL. The GPT-2's embedding size is set to 768, and the sequence length is configured to 1024, allowing for the effective extraction of hidden information from the extended protocols. Subsequent to the application of the LLM, the extracted hidden information is processed through 6 layers of self-attention with 6 attention heads and 6 layers of cross-attention with 6 attention heads. This structure aids in recognizing and processing the comprehensive relationships between the protocols and the associated query entities. Further details of the model's configuration are presented in Table~\ref{tab:configuration}.

\begin{table}[]
\caption{Configuration of Model}
\label{tab:configuration}
\centering
\begin{tabular}{|l|l|}
\hline
\textbf{Parameter}                     &\textbf{Value} \\ \hline
GPT-2's Embedding Size          & $768$   \\ \hline
GPT-2's Sequentce Length        & $104$   \\ \hline
GPT-2's Layers                  & $12$    \\ \hline
GPT-2's Attention Head          & $12$    \\ \hline
Cross-attention Layer          & $6$     \\ \hline
Cross-attention Attention Head & $6$     \\ \hline
Self-attention Layer           & $6$     \\ \hline
Self-attention Attention Head  & $6$     \\ \hline
Learning Rate                  &  $1e^{-7}$     \\ \hline
Epoch                 &  $100$     \\ \hline
dropout rate                 &  $0.1$     \\ \hline
Train/Validation Ratio                &  $9:1$     \\ \hline
\end{tabular}

\end{table}

\subsection{CAL Experiment Result Analysis}\label{sec:result_analysis}

\begin{figure}
    \centering
    \includegraphics[width=0.5\textwidth]{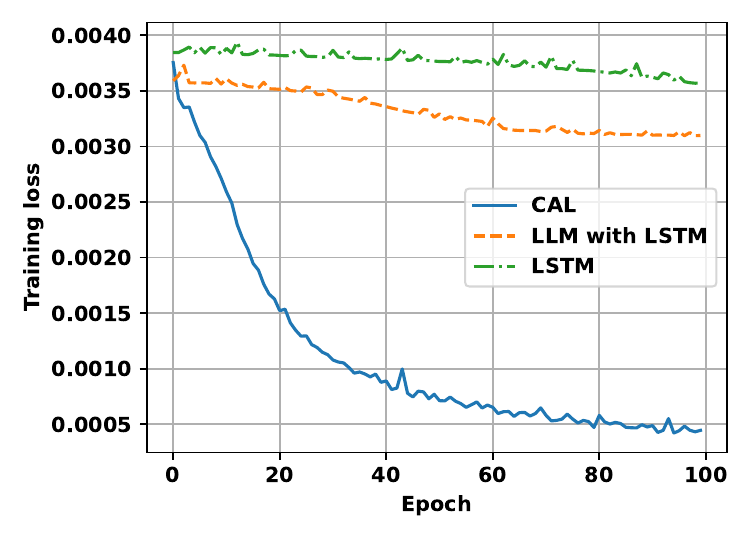}
    \caption{Training Loss: the training loss indicate the incapability of traditional models like LSTM in processing complex dependency and text.}
    \label{fig:training_loss}
\end{figure}

As depicted in Fig.~\ref{fig:training_loss}, the consistent trends in both training and validation accuracy of CAL underscore the model’s proficiency in extracting the dependency relationship information from the examined protocols and entities. In our experiment setting, formal properties are designated as positive, while their absence is categorized as negative. As presented in Table\ref{tab:accuracy}, our model performs stable accuracy of $95.94$\%, and achieves $100$\% accuracy in recall, indicating the high and stable performance in formal property prediction and the trustworthiness of the model. 

\begin{figure}
    \centering
    \includegraphics[width=0.5\textwidth]{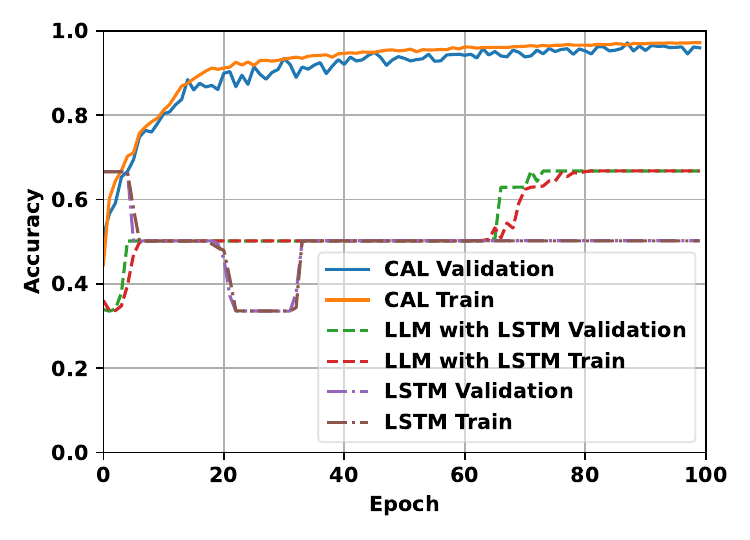}
    \caption{Training Result: Suboptimal models, such as those utilizing LLM with LSTM and standalone LSTM, often get trapped in local minima due to Class Imbalance and Specially Designed Loss.}
    \label{fig:train_valid_accuracy}
\end{figure}

We further compared our model with other state of art model performance.  Table~\ref{tab:accuracy} leads to an explicit conclusion that the CAL significantly outperforms all other models delineated in Section~\ref{experiment_setting}. Fig.~\ref{fig:train_valid_accuracy} reveals that only the LSTM model and the hybrid LLMs with LSTM model exhibit a deficiency in the efficient extraction of information during the training phase, resulting in trapping in local optima. The result also indicates the necessity of LLMs in processing complex protocols and standards. 

To compare the performance of transformer-based models with traditional non-transformer models, the LSTM, known for effectively capturing long-range dependencies and typically performs better than CNN and RNN, is chosen to replace the cross-attention and self-attention. To ensure consistency of experimentation, the LSTM is set with a hidden state size of $768$ and $6$ recurrent layers. Additionally, the LSTM alone was used to directly process natural language to predict formal properties, also with a hidden state size of $768$ and $6$ layers.

\begin{table}[]
\caption{Accuracy of Different Models}
\centering
\label{tab:accuracy}
\begin{tabular}{|l|l|}
\hline
Model         & Accuracy \\ \hline
CAL           & 95.94\%  \\ \hline
LLM with LSTM & 66.76\%  \\ \hline
LSTM          & 50.21\%  \\ \hline
\end{tabular}
\end{table}

\begin{figure}
    \centering
    \includegraphics[width=0.5\textwidth]{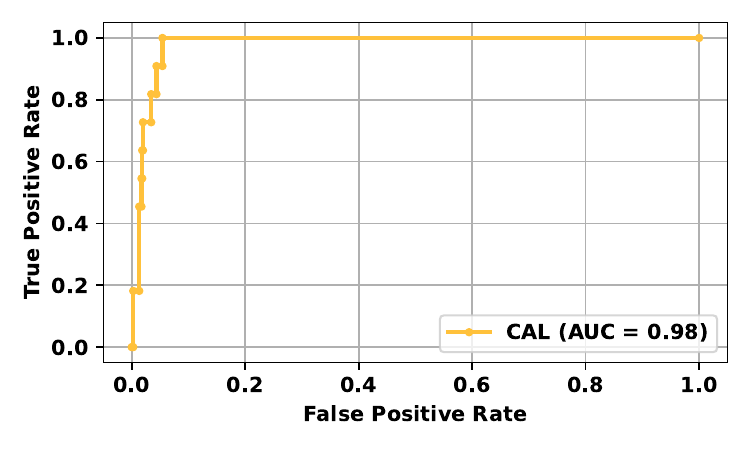}
    \caption{Receiver Operating Characteristic (ROC) curve of CAL Model}
    \label{fig:roc}
\end{figure}

The Area Under the Curve (AUC) is commonly described as a statistical measure used to evaluate the performance of a classification model. As shown in Fig.\ref{fig:roc}, we can see that via the True Positive Rate (TPR) against the False Positive Rate (FPR), CAL has an AUC value of 0.98, indicating excellent performance. 

\begin{figure}
    \centering
    \includegraphics[width=0.5\textwidth]{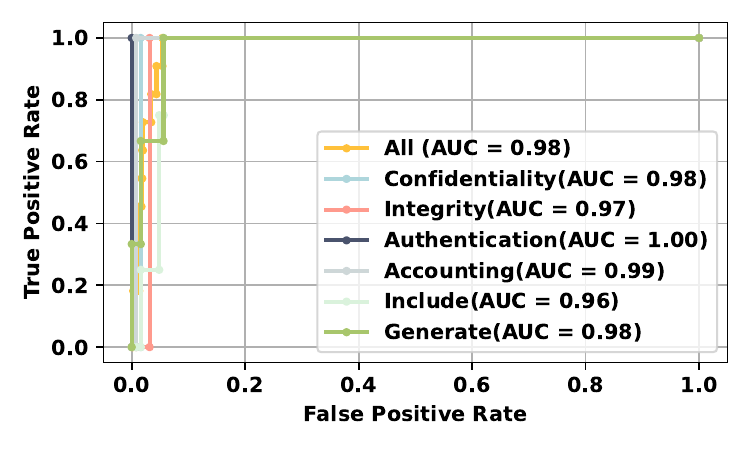}
    \caption{Dependency Relations Specific ROC curve of CAL Model}
    \label{fig:auc_dependency}
\end{figure}

Then, we further analyze the AUC for each type of dependency, as shown in Fig. \ref{fig:auc_dependency}. Confidentiality: This curve appears to be the closest to the top-left corner, suggesting it has the best performance among the four criteria.
Integrity, Authentication, Accounting: These curves are further from the top-left corner compared to Confidentiality. While they still demonstrate good performance, they are not as effective as the Confidentiality curve.

To better visualize the interactions between relevant entities and protocol procedure descriptions, we utilize the Actions related to the transmission of ``RRCSetupRequest'' message as an example. This verification process aims to determine whether our model can effectively extract valuable information from the human-written protocol procedure descriptions.

\begin{figure}[h!]
\centering
    \includegraphics[trim={0.7cm 0 0 0},width=0.48\textwidth]{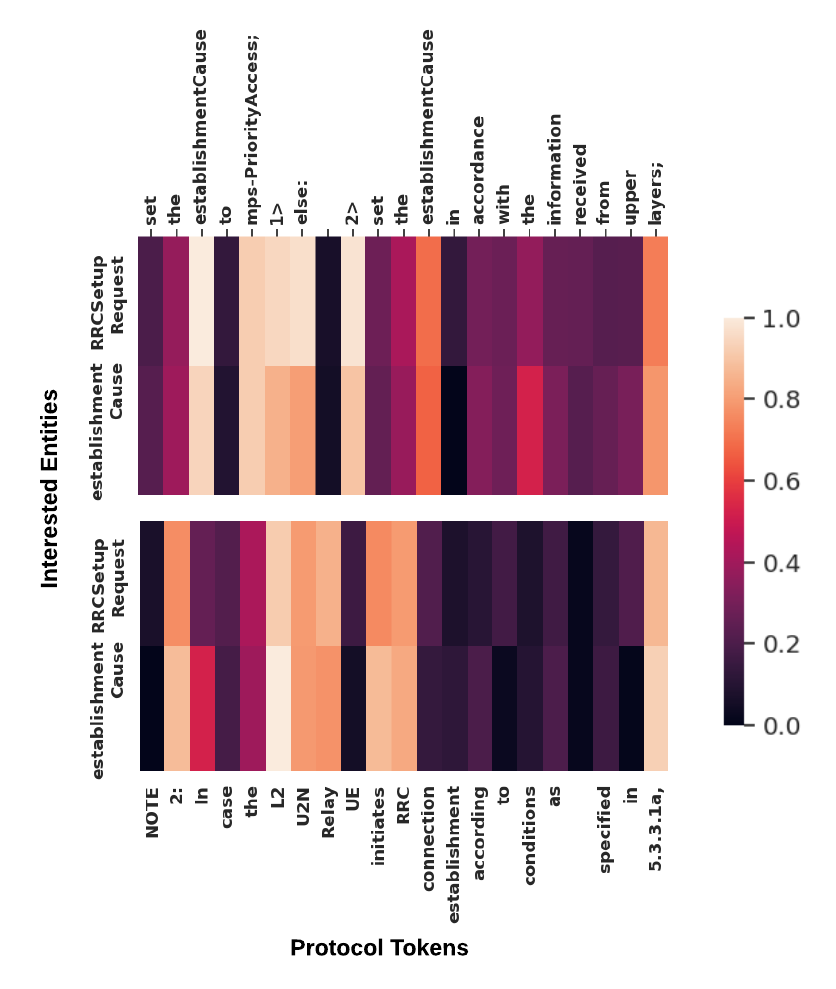}
    \caption{Attention Map Depicting Average Attention Scores Across Layers: Lighter Shades Indicate Higher Attention; Darker Shades Signify Reduced Attention.}
    \label{fig:rrc_attention_map}
\end{figure}

In transformer-based models, each transformer layer has the capability to extract distinct information through different attention scores. In this study, we leverage the average score of cross-attention layers to visualize the interactions between relevant entities and protocol procedure descriptions. Taking the entities ``establishmentCause'' and ``RRCSetupRequest'', along with their corresponding procedure as an example, we generate an attention map (Fig.~\ref{fig:rrc_attention_map}). The attention map clearly reveals that these two entities place more emphasis on specific terminologies, such as ``establishmentCause'', ``mps-PriorityAccess'', and ``RRC'', as well as the levels of indentation, such as ``1$>$'' and ``2:''. This observation highlights the cross-attention component's ability to effectively focus on and extract essential information from key terms within the input text.

\subsection{Case Study of Design Intention Capturing and 
Trustworthy Enhancement via the connection to real-world testbed }

\textbf{Case 1: Design Intention} With the CAL model, the generated dependencies include both intended and unintended ones. Compared to extraction from human expertise, a portion of unintended dependency relationships is also detected. These unintended dependencies can be filtered out via connection to a real-world experimental platform.

\begin{figure*}[h!]
\centering
    \includegraphics[width=\textwidth]{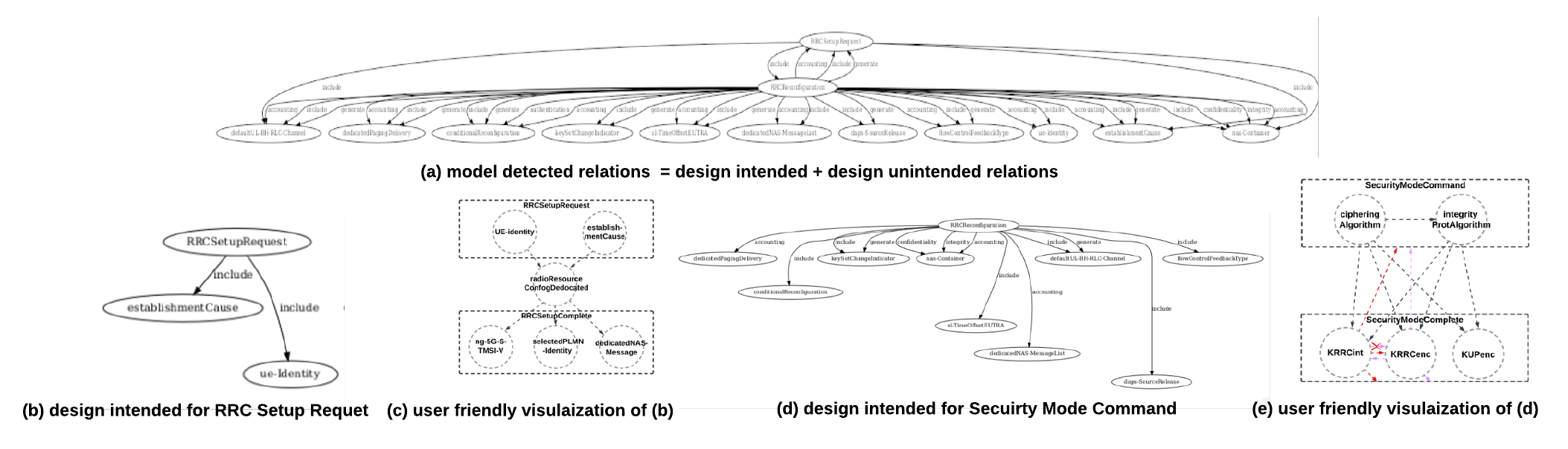}
    \caption{Design Intention: (a) is the raw dependency relationships that detected by CAL, and (b) is the experiments filtered design intended dependency graph.}
    \label{fig:design_intention}
\end{figure*}

\begin{figure}[hbt!]
\centering
    \includegraphics[width=0.5\textwidth]{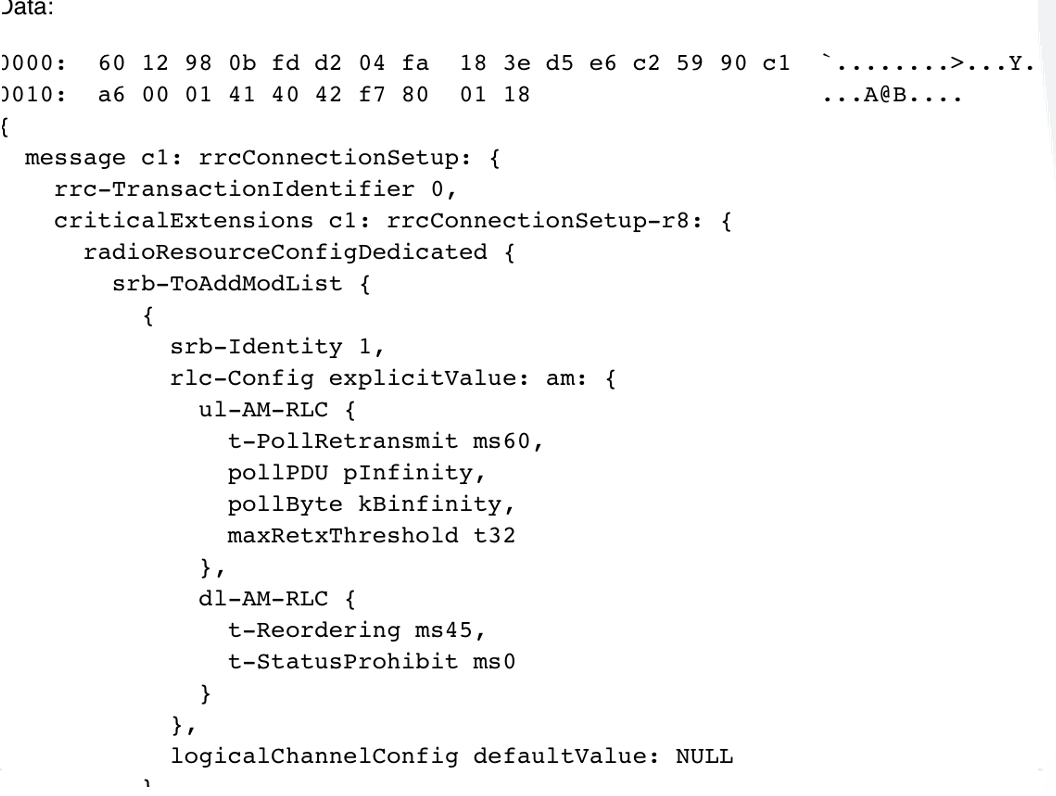}
    \caption{Information Flow-graph detected in Experiment Platform}
    \label{fig:information_flow_graph}
\end{figure}

We took the example of ``RRC\_setup\_request'' and ``Security\_mode\_command'', as shown in Fig. \ref{fig:design_intention}. Figure \ref{fig:design_intention}(a) illustrates the raw dependency relationships detected by CAL, while Figure \ref{fig:design_intention}(b) shows the experimentally filtered, design-intended dependency graph. Using the information flow from the experiment platform, as shown in Fig.\ref{fig:information_flow_graph}, we can derive the design-intended dependency graph presented in Fig. \ref{fig:design_intention}(b) and Fig. \ref{fig:design_intention}(d). Figures \ref{fig:design_intention}(c) and \ref{fig:design_intention}(e) are equivalent to Figures \ref{fig:design_intention}(b) and \ref{fig:design_intention}(d) respectively, offering a more user-friendly visualization. 

\begin{figure}[hbt!]
\centering
    \includegraphics[width=0.5\textwidth]{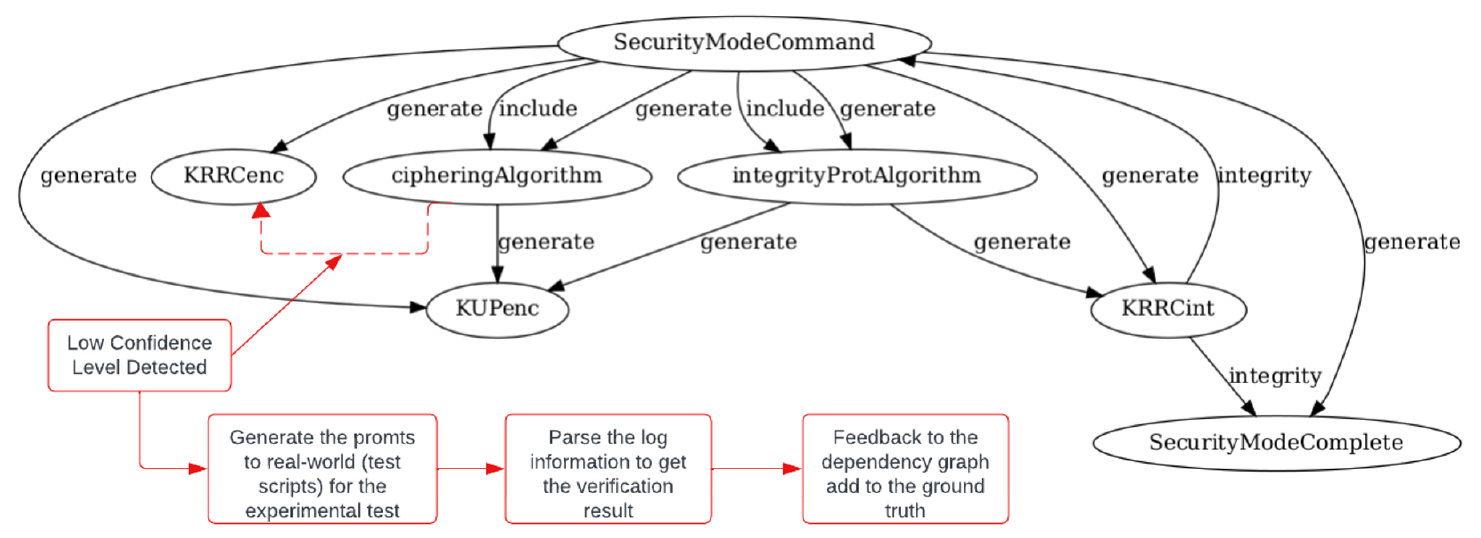}
    \caption{Information Flow-graph detected in Experiment Platform}
    \label{LLM_Trustworth}
\end{figure}

\begin{figure}[hbt!]
\centering
    \includegraphics[width=0.5\textwidth]{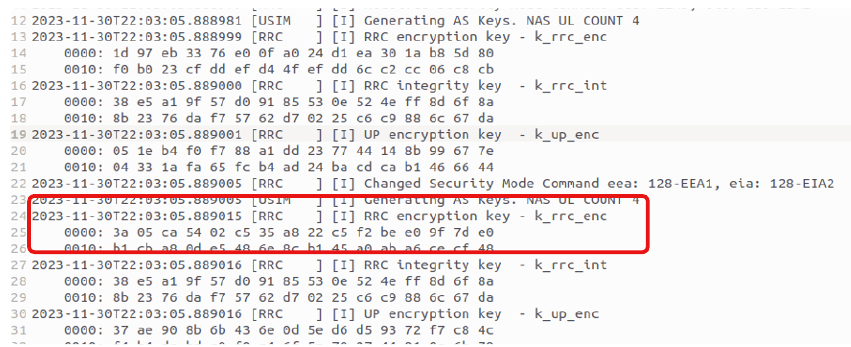}
    \caption{Fuzzing Log File to Parse Evidence Information}
    \label{fig:log}
\end{figure}

\textbf{Case 2: Trustworthy} For dependencies predicted by CAL with low confidence, the uncertain dependency can be converted into test scripts and sent to an experimental platform to generate evidence. This evidence can be used to confirm or refute the existence of the dependency. As shown in Fig. \ref{LLM_Trustworth}, the dependency between KRRCenc and CipherAlgorithm shows low confidence in the detected ``Integrity''. This is automatically converted into fuzzing scripts to generate evidence, proving that there is no dependency between them, as indicated in the parsed information from the log file \ref{fig:log}.

\begin{figure}[h!]
    \centering
     \subfloat[RRC Connection Flow Graph]{\includegraphics[width=0.23\textwidth]{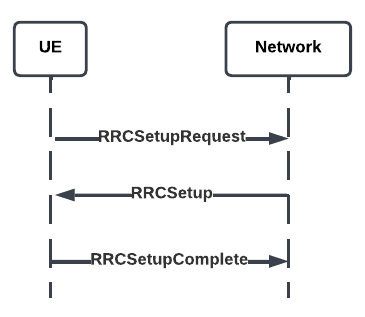}\label{fig:rrc}}
    \subfloat[Security Mode Command Flow Graph]{
                    \includegraphics[width=0.23\textwidth]{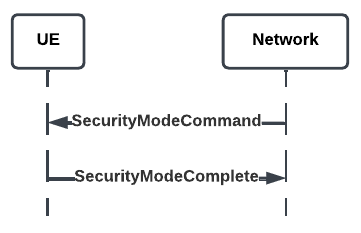}\label{fig:as}}
    \caption{Flow Graph}
    \label{fig:flow}
\end{figure}

\begin{figure}[hbt!]
    \centering
     \subfloat[Predicted RRC Connection Dependency Graph]{\label{fig:rrc_connect}\includegraphics[width=0.23\textwidth]{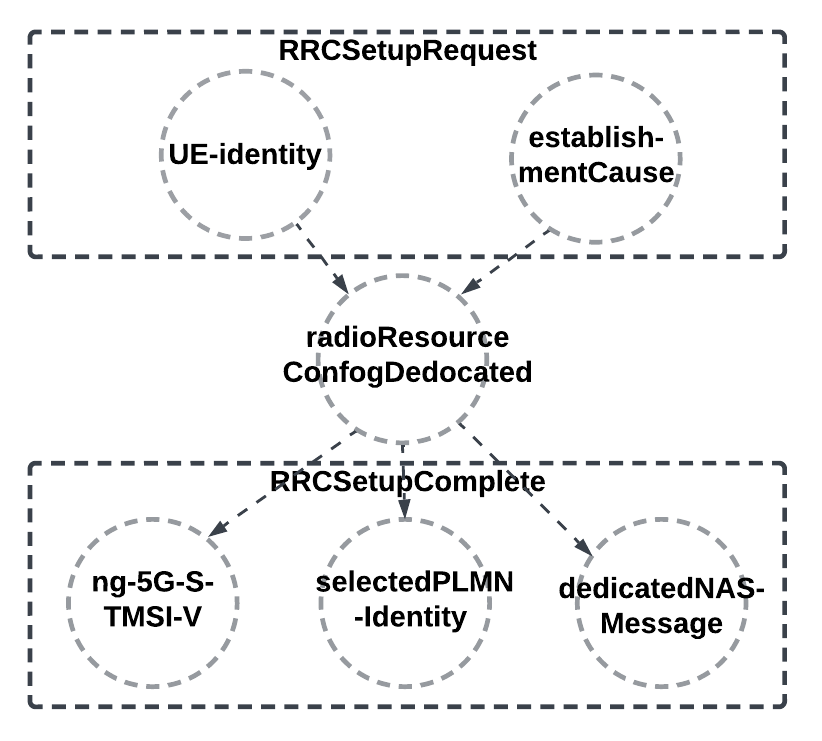}}
      \subfloat[Ground Truth RRC Connection Dependency Graph]{\label{fig:gt_rrc_connect}\includegraphics[width=0.23\textwidth]{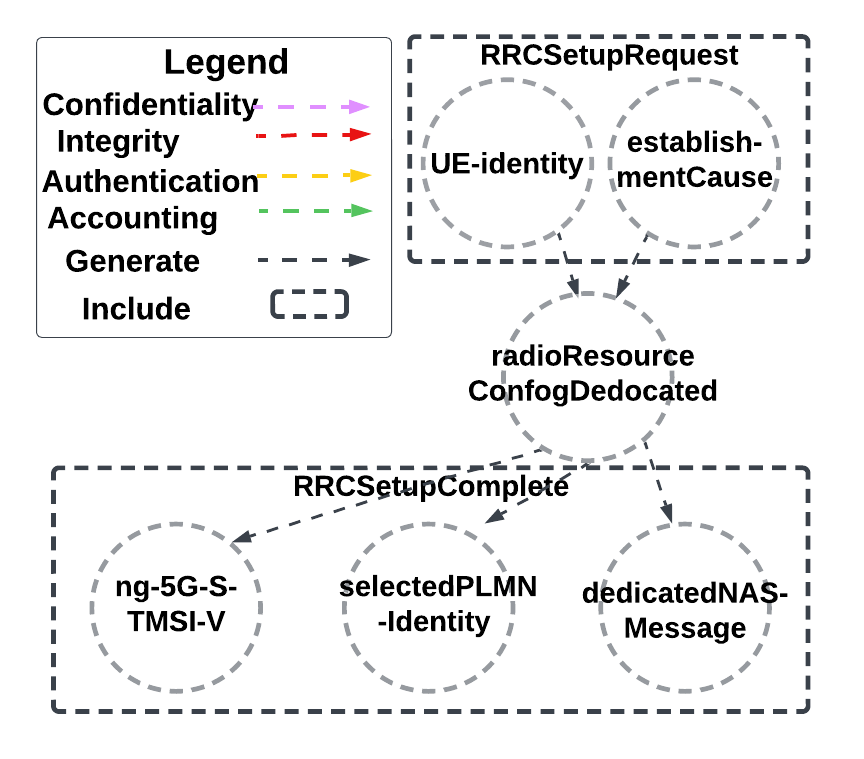}}
          
    \subfloat[Predicted Security Mode Command Dependency Graph]{\label{fig:security_mode}\includegraphics[width=0.23\textwidth]{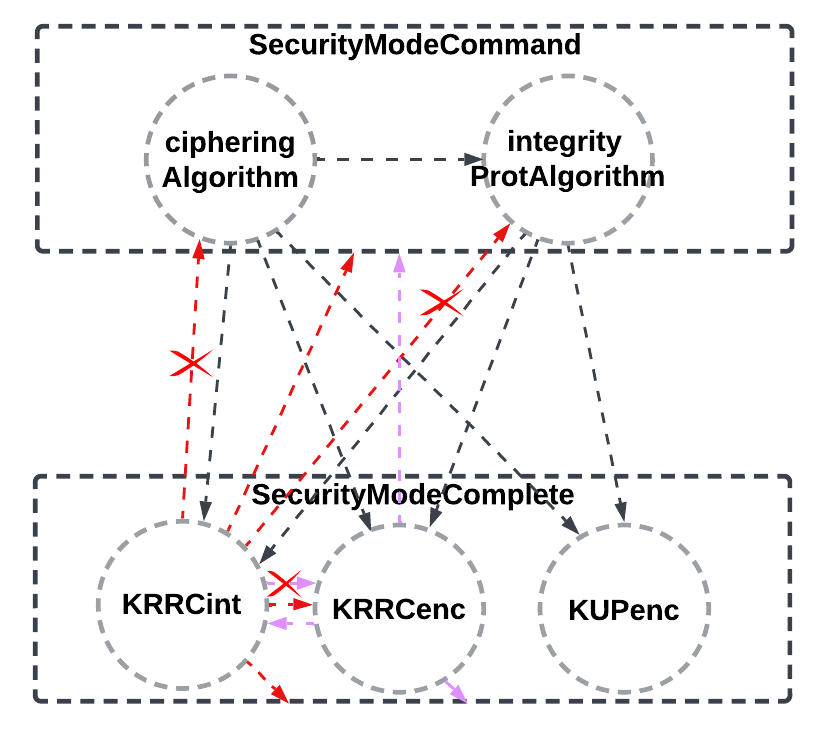}}
    \subfloat[Ground Truth Security Mode Command Dependency Graph]{\label{fig:gt_security_mode}\includegraphics[width=0.23\textwidth]{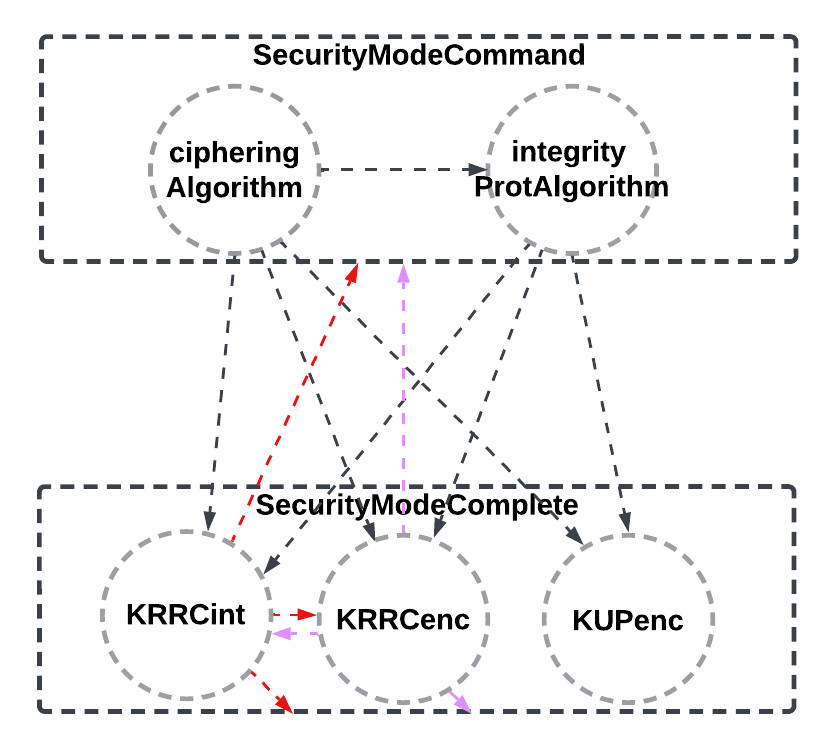}}
    \caption{Comparison between Predicted and Ground Truth Dependency Graph}
    \label{fig:comparison}
\end{figure}

\subsection{Formal Verification and Attack Model}

Based on the predicted relationship between entities, our next objective is to perform the formal verification, as shown in Fig.~\ref{fig:flow}. The interconnected sections and communicated commands consisting of identifiers. We can first fill up the interconnected sections and communicated commands, which is depicted in natural language protocols, shown in Fig.~\ref{fig:rrc} and Fig.~\ref{fig:as}. And from the command table in natural language protocols, we can easily identify which identifiers are included in the commands.

Utilizing the identified formal properties, we construct a comprehensive dependency graph, which facilitates the derivation of formal dependencies through this graph.

We consider the RRC connection establishment procedure as an example. Nodes were manually extracted from the natural language protocol, as depicted in Figs. \ref{fig:rrc_connect} and \ref{fig:gt_rrc_connect}. The visual representation employs boxes to signify ``Include'' and uses various arrow types to delineate distinct formal properties. Leveraging our predicted formal properties, we generated the RRC connection dependency graph (see Fig. \ref{fig:rrc_connect}). Remarkably, this aligns perfectly with the ground truth of the RRC connection dependency graph. However, during the Security Mode procedure, there were discrepancies between the predicted formal properties (as presented in Fig. \ref{fig:security_mode}) and the ground truth (see Fig. \ref{fig:gt_security_mode}). Therefore, subsequent manual verification is advisable. Notably, while there may be the inclusion of extraneous formal properties, none are omitted. Building on previous related work \cite{yang2023formal}, this dependency graph can be transformed into Proverif code, enabling formal verification. Compared to solely manual labeling of formal properties, our proposed model streamlines the process by narrowing down the entirety of natural language protocols to the task of parsing redundant formal properties from the crafted formal dependency graph.







\section{Conclusion}\label{conclusion}

In conclusion, we present AVRE, a novel system for the formal verification of NextG protocols, leveraging Large Language Models (LLMs) to transform protocol descriptions into dependency graphs and formal models. Enhanced by the HyFuzz experimental platform, AVRE demonstrates significant advances in the accuracy and relevance of formal verification in complex communication protocols. The research underscores the efficiency of CAL, a continuously-learning, cross-attention-based LLM, in extracting formal properties and dependencies, outperforming traditional methods. The study emphasizes the potential of LLMs in enhancing trustworthiness and clarifying ambiguities in protocol verification, marking a significant contribution to the field. By reducing reliance on manual labeling and associated human errors, our method offers a more efficient approach, focusing only on the most pertinent formal relationships. Our work in extracting formal relationships from natural language protocols enhances the clarity and understanding of these protocols, ensuring a more robust, reliable, and efficient approach for protocol verification and system validation in large-scale, complex systems.




\section*{Acknowledgment}
This effort was sponsored by the Defense Advanced Research Project Agency (DARPA) under grant no. D22AP00144. The views and conclusions contained herein are those of the authors and should not be interpreted as necessarily representing the official policies or endorsements, either expressed or implied, of DARPA or the U.S. Government.

\bibliographystyle{IEEEtran}
\bibliography{Bibliography,sample}

\end{document}